# Stress-induced martensitic transformation in epitaxial Ni-Mn-Ga thin films and its correlation to optical and magneto-optical properties

M. Makeš[1, 2, *], J. Zázvorka[1], M. Hubert[1], P. Veřtát[2], M. Rameš[2], J. Zemen[2], O. Heczko[2] and M. Veis[1, †]

[1] *Faculty of Mathematics and Physics, Charles University, Ke Karlovu 3, 12116 Prague 2, Czechia*

[2] *FZU – Institute of Physics of the Czech Academy of Sciences, Na Slovance 1999/2, 18200 Prague 8, Czechia*

[*] Contact author: makes@fzu.cz

[†] Contact author: martin.veis@matfyz.cuni.cz

The optical and magneto-optical properties of thin epitaxial Ni-Mn-Ga films, with thicknesses ranging from 8 to 160 nm, were investigated across the spectral range of 0.7–6.4 eV. The films were deposited by DC magnetron sputtering on MgO substrate with stress-mediating Cr buffer layer. Structural and magnetic characterization revealed a stress-induced martensitic transformation for the films thicker than 80 nm, while thinner films remained in austenite structure due to the substrate constraint. In martensitic samples, a mixture of 14M and NM phases was observed with both X-type and Y-type twin domains. Magneto-optical polar Kerr effect spectra showed notable evolution with film thicknesses, demonstrating changes in electronic structure of Ni-Mn-Ga. A combination of spectroscopic ellipsometry and magneto-optical Kerr spectroscopy allowed for the deduction of the spectral dependence of full permittivity tensor of investigated samples. Three magneto-optical transitions were fitted to the spectra of non-diagonal permittivity using semiclassical theory, showing strong correlations with in-plane coercivity. Observed correlations highlight the impact of substrate-induced stress on Ni-Mn-Ga films and provide insights into the electronic structure changes associated with the martensitic transformation.

# I. INTRODUCTION

The shape-memory effect (SME) originates from the thermoelasticity of the martensitic transformation (MT), a diffusionless displacive transition from a high-temperature cubic austenitic phase to a low-temperature martensitic phase. Cooling the shape-memory alloy (SMA) into the martensitic phase and applying mechanical deformation allows restoration of its original shape by heating, resulting in shape memory behavior [1]. The magnetic-shape memory effect (MSME) is driven by magnetic field [2], either by inducing the martensitic transformation [3] or by magnetically induced reorientation (MIR) of twin variants in the martensitic phase [4]. This enables much higher switching frequencies compared to conventional SMA, which relies on temperature changes.

Since the first observation of MSME (0.2% deformation) on stoichiometric $Ni_2MnGa$ single crystal by Ullakko *et al.* [4], large macroscopic deformation (5.1%) due to the MIR was reported for off-stochiometric $Ni_{48}Mn_{31}Ga_{21}$ single crystal [5]. The largest reported deformation due to the MIR was 12% in Co and Cu doped off-stoichiometric Ni-Mn-Ga single crystal [6].

From an applications perspective, magnetic shape memory alloys (MSMAs) in the form of thin films are highly attractive for integration into micromechanical devices such as microactuators [7,8] or micropumps [9]. Epitaxial films offer a well-controlled model system with high crystalline quality, smooth surfaces, and well-defined crystallographic orientation, making them not only promising for device miniaturization but also ideal for fundamental studies of structure–property relationships by optical means. Epitaxial growth of Ni-Mn-Ga films was demonstrated on various substrates, including GaAs(001) with $Sc_{0.3}Er_{0.7}As$ buffer layer [10], $Al_2O_3$(110) [11], NaCl(001) [12,13], Si(100) [14], Y:$ZrO_2$(001) [15], and $SrTiO_3$(001) [16]. However, the most extensive research was focused on MgO(001) [17–25] and MgO(001) with Cr buffer layer [8,26–30]. In both cases the epitaxial relationships MgO(001)[001]‖Ni-Mn-Ga(100)[011] and MgO(001)[001]‖Cr(100)[011]‖Ni-Mn-Ga(100)[011], respectively, were observed. For the preparation of freestanding films, various methods were employed, such as focused ion beam (FIB) etching of the MgO(001) substrate [25], dissolution of the NaCl(001) substrate [12,13], or chemical etching of the Cr buffer layer on the MgO(001) substrate [26]. Different martensite microstructures were observed on free standing films [31]. Although the 10M martensite was observed on epitaxial films grown on NaCl(001) substrate [12,13], for all other substrates (and corresponding free-standing films), only NM and 14M phase were observed.

Kaufmann *et al.* [32] explained that in thin films, the rigid substrate constraint stabilizes thermodynamically unfavorable phases, such as 14M martensite, in contrast to the 10M phase preferred in bulk materials. According to the hierarchical nomenclature of Diestel *et al.* [28], nanotwinning (modulation) is followed by mesoscopic (second generation) twinning. Using the substrate reference frame, the 6 symmetrically equivalent $\{110\}_{MgO}$ twin boundary (TB) planes can be differentiated in the presence of the substrate between two types of mesoscopic *a*-*c* twinning microstructures. X-type twins have TB planes inclined 45° to the surface normal, resulting in surface corrugation, while Y-type twins have TBs perpendicular to the surface normal without surface corrugation.

Magnetic properties of 14M martensite epitaxial films were reported to be similar to their bulk counterparts [17–19,21,29,33]. The effect of substrate constraint was observed at magnetic and martensitic transformation temperatures. Chernenko *et al.* [34] reported about 21 K increase of Curie temperature between 30 nm and 1000 nm films. Ranzieri *et al.* [35] studied the influence of film thickness on structural, magnetic, and surface properties of epitaxial $Ni_{52.5(9)}Mn_{19.5(7)}Ga_{28.0(5)}$ films deposited onto MgO(001) substrate at 420 °C. Thinner films (10 and 20 nm) remained in the austenitic phase upon cooling from the deposition temperature across the measured temperature range, exhibiting flat island-like surface morphology with no observable magnetic-force-microscopy (MFM) contrast. Conversely, thicker films (40, 75 and 100 nm) underwent MT with 6–16 K temperature hysteresis, with coercivity reported in the 2–14 mT range.

Thomas *et al.* [17] measured substrate-induced tetragonal distortion and estimated biaxial tensile stress of about 100 MPa, causing a tetragonal distortion of cubic lattice in epitaxial 473 nm $Ni_{52.0(5)}Mn_{23.0(5)}Ga_{25.0(5)}$ film deposited directly on MgO(001) substrate. Biaxial symmetry does not allow shear stress in the $(001)_{NMG}$ plane, thus they observed only X-type twins with TBs inclined 45° to the surface normal and intersecting the film surface in the $<110>_{NMG}$ directions [19,36].

Examination of changes in electronic structure associated with MT is crucial for understanding its origin and improving desired material properties. Spectroscopic methods, including photoemission spectroscopy, spectroscopic ellipsometry (SE) and generalized magneto-optical spectroscopic ellipsometry, are suitable for this task. Phase transitions were studied in various materials using these methods, including $Pb_{1-x}Ge_xTe$ [37], $HfO_2$ films [38], Zr [39], transition metal dichalkogenide single crystals [40], Fe-Rh films [41], Ni-Ti [42,43], and Heusler alloys [44–59].

While several studies [44,45,54,60,61] on Ni-Mn-Ga single crystals and thin films have employed magneto-optical spectroscopy in various geometries, the influence of the substrate constraint and film thickness has not yet been systematically investigated using this technique. Moreover, as the differently thick films exhibit different phases, the effect of MT on changes in optical and magnetooptical properties can be revealed, allowing for analysis of electronic structure changes upon MT. Here, we report about systematic study of the evolution of optical and magneto-optical properties of Ni-Mn-Ga thin films with increasing thickness. The stress-induced MT was observed and its fingerprints in optical and magneto-optical spectra were described. Using semiclassical theory of magneto-optical transitions, we were able to find correlations between fitted parameters and magnetic properties of investigated films.

# II. EXPERIMENTAL

Investigated epitaxial Ni-Mn-Ga films with nominal thicknesses of 10, 20, 50, 100, and 200 nm were deposited by DC magnetron sputtering in high vacuum ($10^{-9}$ Pa) at 500 °C on nominal 20 nm buffer Cr(100) layer and MgO(100) substrate using $Ni_{50}Mn_{25}Ga_{25}$ polycrystalline target. The layers grow with their unit cell orientations MgO(001)[001]‖Cr(100)[011]‖Ni-Mn-Ga(100)[011], as commonly observed [26–30]. The twin microstructure, layer interfaces, and layer thicknesses were observed by transmission electron microscopy (TEM) using FIB-made lamella cut from the 100 nm film in the $[101]_{MgO}$ orientation. Surface morphology was observed by atomic force microscopy (AFM) Bruker. Composition was determined by X-ray fluorescence (XRF) and valence electron concentration per atom *e*/*a* was calculated from atomic percentages (see **Table 1**). Since all films were prepared in the same conditions with the same target and only deposition time was varying, it can be assumed that the films have the same composition. The atomic percentages vary mainly due to decreasing XRF precision thinner films. The compositional deviation (decrease of Mn content in favor of Ni and Ga) from the target is common during sputtering because of high volatility of Mn.

Table 1: Stoichiometric compositions, corresponding valence electron concentration *e*/*a* and thicknesses of studied thin Ni-Mn-Ga films. * TEM showed the nominal 100 nm thick NMG film to be 76 nm thick with 28 nm Cr layer, which was used as a fixed parameter for XRR analysis of other films. ** The thickness of the thickest film could not be determined by XRR, but conformity to the systematic 20 %, decrease from the nominal thickness was assumed.

| Nominal thickness (nm) | Ni (at. %) | Mn (at. %) | Ga (at. %) | *e*/*a* | XRR thickness (nm) |
|---|---|---|---|---|---|
| 10 | 50(3) | 17(2) | 33(4) | 7.2(4) | 8(4) |
| 20 | 51(2) | 17(1) | 32(2) | 7.3(2) | 16(4) |
| 50 | 50(1) | 19(1) | 31(1) | 7.3(1) | 38(4) |
| 100 | 51.4(7) | 19.0(4) | 29.6(8) | 7.36(8) | 80(4)* |
| 200 | 51.3(3) | 19.5(5) | 29.2(6) | 7.37(5) | 160** |
| average | 51(1) | 19(1) | 30(1) | 7.3(1) | 60 |
| bulk | 51.75(4) | 19.49(7) | 28.76(7) | 7.40(1) | — |

X-ray diffraction (XRD) experiments were carried out using two diffractometers: PANalytical X'Pert PRO with Co anode ($\lambda_{K\alpha 1}$ = 0.17890 nm), which was used mainly for preliminary scans, pole figures, simple 1D scans in parallel beam geometry, and Rigaku SmartLab with Cu anode ($\lambda_{K\alpha 1}$ = 0.15406 nm) in parallel beam geometry and HyPix-3000 area detector, which was mainly used for reciprocal space mapping.

The thickness of the films and buffer layer was determined by X-ray reflectivity (XRR) performed using the Rigaku SmartLab diffractometer. A double-bounce Ge(220) monochromator was used for the 160 nm film, while the rest of the films were measured without monochromator. Theoretical XRR curve was fitted to the data using the software package GenX 3 (see **Fig. S7** in the Supplementary Material [62]), with a fixed thickness of the Cr layer (assumed to be 28(4) nm for all samples). According to the XRR results, all samples have thicknesses approximately 20% lower than the nominal value.

A reference polycrystalline bulk $Ni_{52}Mn_{19}Ga_{29}$ was prepared by arc melting high-purity elements (≥99.9%) in an argon overpressure atmosphere using a MAM-1 furnace (Edmund Bühler GmbH, Bodelshausen, Germany). Samples measuring approximately 5 mm × 3 mm × 1 mm were cut from the annealed portion using a spark erosion machine (ZAP BP) and roughly polished with SiC grinding paper up to a 2400 grit.

Magnetic measurements were performed using the vibrating sample magnetometer (VSM) option of the system PPMS (Quantum Design). Magnetic field was oriented in-plane along the $[100]_{MgO}$ direction. The temperature dependence of magnetic moment in the magnetic field $\mu_0 H = 0.01$ T was measured at the 4 K/min rate, starting with heating from 300 K up to 400 K, followed by cooling down to 10 K, heating up to 400 K, and finishing with cooling back to 300 K. Magnetic hysteresis loops were measured at fields between −9 and +9 T at rates of 1 and 20 mT/s, the former one being used between −0.1 and 0.1 T in order to get detailed information about magnetic hysteresis. Precise coercivity measurements were conducted using Microsense VSM along the $[100]_{MgO}$, $[110]_{MgO}$, and $[001]_{MgO}$ directions at room temperature (RT) in magnetic field up to 1.8 T.

Optical measurements of the relative permittivity tensor $\varepsilon_{xx}$ were done at RT in the reflection setting for 6 different angles of incidence equidistantly spread from 45° to 70° in the 0.73–6.42 eV spectral range by J.A.Woollam RC2 ellipsometer with two rotating compensators. Polar magneto-optical Kerr effect (PMOKE) was measured with rotating analyzer ellipsometer for near-normal incidence in the 0.76–5 eV spectral range at RT. The sample was mounted on a pole of an electromagnet, generating 1 T magnetic field oriented parallel to the normal of the sample surface, producing nearly magnetically saturated state. To eliminate magnetization-even effects, spectra were averaged over both magnetic field polarities and 3 iterative cycles yielded precision in the 0.001°–0.003° range.

In this work, we follow the sign convention, in which field quantities propagate with the factor $e^{i(\omega t-\boldsymbol{k}\cdot\boldsymbol{r})}$, leading to negative imaginary part of response tensor, i.e., permittivity $\varepsilon_{ij} = \mathrm{Re}\,\varepsilon_{ij} - i\mathrm{Im}\,\varepsilon_{ij}$. Linear optical and magneto-optical (MO) responses of originally isotropic film in polar geometry can be described by the relative permittivity tensor [63]

$$\boldsymbol{\varepsilon} \approx \begin{pmatrix} \varepsilon_{xx} & \varepsilon_{xy} & 0 \\ -\varepsilon_{xy} & \varepsilon_{xx} & 0 \\ 0 & 0 & \varepsilon_{xx} \end{pmatrix}, \tag{1}$$

where the z-axis is parallel both to the surface normal and the external magnetic field. From time inversion symmetry, it follows that $\varepsilon_{xx}$ is even in magnetization, while $\varepsilon_{xy}$ is odd. Thus, the linear MO response will manifest only in $\varepsilon_{xy}$. An alternative description suitable for metals is provided by the conductivity tensor

$$\sigma_{ij} = \mathrm{Re}\,\sigma_{ij} - i\mathrm{Im}\,\sigma_{ij} = i\frac{E}{\hbar}\varepsilon_0(\varepsilon_{ij} - \delta_{ij}), \tag{2}$$

where $E$ is the incident photon energy, $\delta_{ij}$ is the Kronecker delta, $\hbar$ is Planck's constant, and $\varepsilon_0$ is the vacuum permittivity.

Spectral dependences of diagonal elements ($\varepsilon_{xx}$) of the permittivity tensor were obtained by fitting the theoretical model to experimental ellipsometric data $\Psi$ and $\Delta$ (for details see the Supplementary Material [62]). The model consisted of an ideal layered structure (homogeneous layers with plan-parallel boundaries) with no interface roughness and fixed thicknesses as determined by XRR. Tabulated data was used as optical constants of chromium buffer layer [64] and MgO substrate [65]. The optical response of the Ni-Mn-Ga layer was modelled as

$$\varepsilon_{xx}(E) = \varepsilon_\infty + \frac{-\hbar^2}{\varepsilon_0\rho(\tau E^2 - i\hbar E)} + \sum_{i=1}^{3}\frac{A_i G_i E_i}{E_i^2 - E^2 + iEG_i}, \tag{3}$$

i.e., a sum of a real constant $\varepsilon_\infty$, accounting for inter-band absorption at higher energies, a Drude term [66], described by resistivity $\rho$ and scattering time $\tau$, and three damped Lorentzians [67], each characterized by amplitude $A$, energy $E$, and width $G$.

From the measured diagonal permittivity $\varepsilon_{xx}$ and PMOKE spectra, the off-diagonal elements of the relative permittivity tensor $\varepsilon_{xy}$ were obtained point by point from regression analysis in the 0.76–4.6 eV photon energy range using the Yeh $4 \times 4$ matrix formalism [68,69]. According to the semiclassical theory [63], the fitting function

$$\varepsilon_{xy}(E) = \sum_{t}^{\mathrm{para}} G_t A_t \frac{-2G_t(E_t^2 + E^2 - G_t^2) - 2iE(E_t^2 - E^2 - G_t^2)}{((E_t - E)^2 + G_t^2)^2 + 4E^2G_t^2}$$
$$+ \sum_{t}^{\mathrm{dia}} G_t A_t \frac{2G_t(E_t - E) - i((E_t - E)^2 - G_t^2)}{((E_t - E)^2 - G_t^2)^2 + 4G_t^2(E_t - E)^2} \tag{4}$$

for the calculated spectra is composed of paramagnetic transitions (the first sum) and diamagnetic transitions (the second sum), assuming negligible spin-orbital interaction with respect to the transition width. Each magneto-optical transition is determined by its energy $E_t$, half-width at half-maximum (HWHM) $G_t$, and the third parameter

$$A_t = \mathrm{Im}\{\varepsilon_{xx}(E_t)\} \times \begin{cases} \xi_t & \text{for paramagnetic transition} \\ S_t & \text{for diamagnetic transition} \end{cases} \tag{5}$$

is proportional to either the fractional dichroism $\xi_t$ (dimensionless) or the spin-orbit splitting $S_t$ (with the dimension of energy). The fit was achieved by minimizing the sum of off-diagonal conductivity square roots. Note that, the sign convention used by Kahn *et al.* [63], yields Hermitian conjugates of these formulae.

The Vienna *Ab initio* Simulation Package (VASP, ver. 5.4.4) [70,71] was used for calculations of magnetic moments, densities of states (DOS) as well as optical properties [72] of stoichiometric $Ni_2MnGa$. We adopted the generalized gradient approximation (GGA) of Perdew-Burke-Ernzerhof (PBE) functional [73] for the exchange-correlation functional with $U = 1.8$ eV at Mn sites, which leads to correct prediction for elastic constants [74].

# III. RESULTS AND DISCUSSION

## A. Structure and microstructure

AFM micrographs of the surface of 8 nm and 160 nm thick layers are shown in **Fig. 1**. Thinner films (8 nm and 16 nm) appear smooth without visible spatial periodicity (see **Fig. 1 a**). In thicker films (80 and 160 nm), areas with periodic corrugation in the $[110]_{\mathrm{MgO}}$ and $[1\bar{1}0]_{\mathrm{MgO}}$ directions indicate X-type twins (see **Fig. 1 b**), while areas without surface corrugation indicate Y-type twins. The twinning planes of Y-type twins oriented 90° relative to the film plane can be seen in the TEM image of the 80 nm film (see **Fig. S8** in the Supplementary Material [62]).

Room temperature XRD $\theta$–$2\theta$ scans of the investigated samples in the out-of-plane direction (see **Fig. S5** in the Supplementary Material [62]) were used to derive out-of-plane lattice parameters of the films and Cr underlayer (listed in **Table 2**).

For the thinner films (8, 16 and 38 nm thick), the diffraction pattern consisted of a single reflection exactly at the out-of-plane orientation without additional spots or twinning, confirming the austenitic structure. The stress-mediating Cr layer has a cubic lattice parameter of 0.288 nm, resulting in a lattice mismatch of about −1%, which induces biaxial in-plane compressive stress on the Ni-Mn-Ga layer and leads to a distortion of the originating austenite lattice. This gradual in-plane compression of the austenitic cubic cell with increasing thickness is apparent from the decrease of in-plane lattice parameters (also listed in **Table 2**), which were calculated for deformation conserving unit cell volume.

In contrast, the 80 and 160 nm thick films were in a martensitic state. Their complex diffraction patterns can be successfully indexed with a primary 14M martensite phase together with a contribution from non-modulated (NM) martensite. They exhibit the typical twinned microstructure, which is confirmed by pole figure measurements. As an example, the pole figure of the (004) reflection of 14M martensite for the 160 nm film (**Fig. 2**) reveals four orientations spread symmetrically around the out-of-plane direction. Further evidence for the 14M twin microstructure is provided by additional XRD scans and pole figure measurements (see **Figs. S3** and **S4** in the Supplementary Material [62]). For the 160 nm thick film, the approximate martensite lattice parameters were determined to be $a \approx 0.619$ nm, $b \approx 0.580$ nm, $c \approx 0.548$ nm, $\gamma \approx 91°$ for the 14M phase, and $a = b \approx 0.548$ nm, $c \approx 0.661$ nm for the proposed NM phase (see the Supplementary Material [62] for additional details regarding the NM presence).

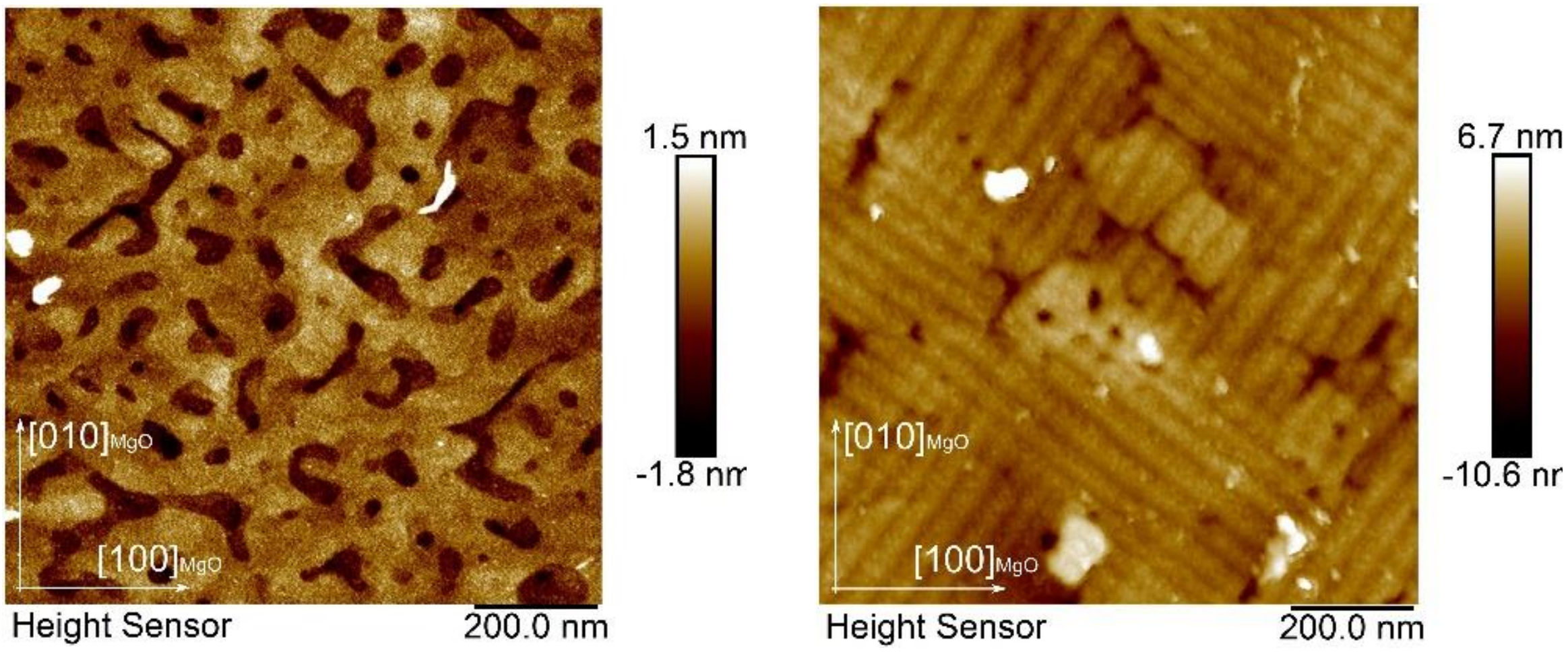

(a) (b)

Fig. 1: (a) AFM image of 8 nm austenitic film, (b) 160 nm martensitic film showing X-type twinning with surface corrugation along the $[110]_{\mathrm{MgO}}$ and the $[1\bar{1}0]_{\mathrm{MgO}}$ directions with the average period of 45 nm.

Table 2: Out-of-plane lattice parameters of Ni-Mn-Ga films and corresponding Cr buffer layer. † In-plane lattice parameters were estimated for tetragonal cell conserving volume of reference bulk cubic cell with the measured lattice parameter included in the table. Estimated in-plane parameters of austenitic (8, 16, and 38 nm) films show increase of tetragonal distortion with increasing thickness. * TEM showed the nominal 100 nm thick NMG film to be 76 nm thick with 28 nm Cr layer, which was used as a fixed parameter for XRR analysis of other films. ** The thickness of the thickest film could not be determined by XRR due to insufficient resolution, but conformity to the systematic 20 %, decrease from the nominal thickness was assumed.

| Film thickness (nm) | Out-of-plane lattice parameter of NMG (nm) | † In-plane lattice parameter of NMG (nm) | Out-of-plane lattice parameter of Cr (nm) |
|---|---|---|---|
| 8(4) | 0.5835(2) | 0.581(1) | 0.2881(2) |
| 16(4) | 0.5843(2) | 0.580(1) | 0.2880(2) |
| 38(4) | 0.5896(2) | 0.578(1) | 0.2880(2) |
| 80(4)* | martensite | martensite | 0.2880(2) |
| 160** | martensite | martensite | 0.2883(2) |
| bulk | 0.5817(5) | cubic | — |

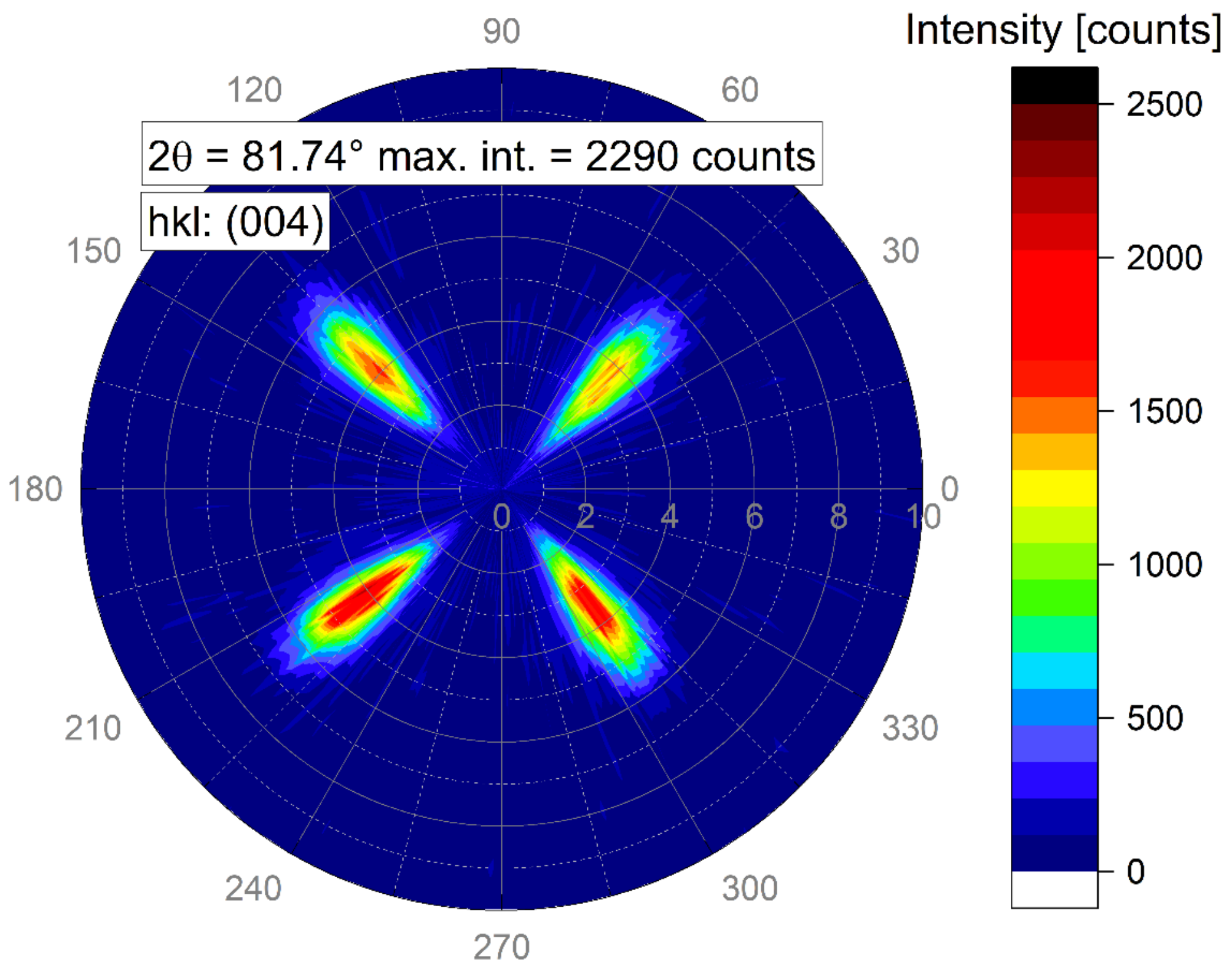

Fig. 2: Zoomed-in detail of the central region ($\chi$ goes from 0° to 10° in this figure) of the pole figure of the $(004)_{14M}$ reflection of the 160 nm thick film. Four orientations spread symmetrically around the out-of-plane direction were detected.

## B. Magnetic properties

To investigate transformation properties of thin film samples, low-field (0.01 T) thermomagnetic curves were analyzed in the 10–400 K temperature range (**Fig. 3 a**). Only the 80 nm and 160 nm thick films underwent martensitic transformation with wide hysteresis in the 280–340 K region near the Curie point. Thinner films (below 40 nm) remained austenitic down to 10 K in agreement with the observations of Ranzieri *et al.* [23]. Moreover, the stress-free reference bulk sample of similar composition did not transform indicating that MT was substrate-induced. Experimental data [75] show martensitic transformations only for alloys with e/a above 7.4, in agreement with this conclusion. It seems that the higher surface-to-volume ratio of thinner samples that are possibly not fully continuous, allows for austenite to be relaxed (see **Fig. 1 a**), while for thicker films this relaxation is inhibited, resulting in stress-induced martensite. Reduced dimensionality alone can suppress martensitic transformation [76–78] as the small thickness does not allow nucleation of sufficiently large nucleation centers for the first order phase transition. Furthermore, using TEM, Ge *et al.* [79] demonstrated a suppression of transformation to 14M martensite on $Ni_{50}Mn_{25}Ga_{20}Fe_5$ single crystal foil in regions thinner than 80 nm. This is in direct agreement with our data.

The Curie temperature $T_C$ was determined from the inflection point on thermomagnetic curve in the region where magnetization falls to zero. With increasing film thickness, $T_C$ linearly rises from 327 to 357 K (blue data in **Fig. 4 a**) in agreement with previous studies [23,34] (excluding the 100 nm outlier reported by Ranzieri *et al.* [23]) and remains well above the reference PC bulk value of 300 K.

In-plane magnetization loops along $[110]_{MgO}$ were measured in 10–400 K range (the exemplary loop is in **Fig. 3 b**, others in the Supplementary Material [62]). The saturated magnetization at $T = 0$ K was estimated using the implicit formula No. 8 in Ref. [80] to be $M_S(0) = 60.5(7)\ \mathrm{Am^2kg^{-1}}$ (theoretical Ni-Mn-Ga density $\rho = 8100\ \mathrm{kgm^{-3}}$ was used). The reason for its value being lower than that of reference PC bulk ($M_S(0) = 69\ \mathrm{Am^2kg^{-1}}$) likely originates from overestimating film volumes due to thickness error and surface roughness.

For precise coercivity evaluation, magnetization loops at RT were measured along the $[110]_{MgO}$, $[100]_{MgO}$ and $[001]_{MgO}$ directions using VSM with electromagnet (exemplary loop can be seen in **Fig. 3 c**). In-plane coercivity at RT varies from 4 to 28 mT ($[110]_{MgO}$, $[100]_{MgO}$ loops differ negligibly). For austenitic samples (8 nm, 16 nm, and 38 nm) with increasing thickness, in-plane coercivity linearly increases, while the out-of-plane coercivity remains constant (8 mT, see **Fig. 4 b**). This well corresponds to compressive stress induced easy axis in-plane and thus coercivity.

For martensitic samples (80 nm and 160 nm), out-of-plane coercivity (24 mT and 28 mT, respectively) is almost twice as high as in-plane coercivity and roughly 4 times higher than the austenitic value. The increase of out-of-plane coercivity in martensite is due to the emergence of complex martensitic microstructure (mixture of NM and 14M, differently oriented 14M mesoscopic twinning—X and Y twins, see **Fig. S8** in the Supplementary Material [62]). Out-of-plane coercivity exceeding in-plane coercivity was also reported for thicker (312–640 nm) Ni-Mn-Ga films on MgO substrate [24]. This is likely due to a change in the magnetic domain structure between austenite and martensite. The maximum in-plane coercivity and the most pronounced Hopkinson peak, indicating high *stress-induced* anisotropy (due to the substrate constraint), were found for the thickest (38 nm) austenitic film, which has the highest tetragonal distortion (see **Table 2**).

Field dependent PMOKE and RT out-of-plane magnetization loops had identical shapes both for the 8 nm and the 160 nm samples (see **Fig. S12 c, d** in the Supplementary Material [62]), confirming the same surface and volume behavior for both austenite and martensite phases in contrast to similar measurement on bulk [48].

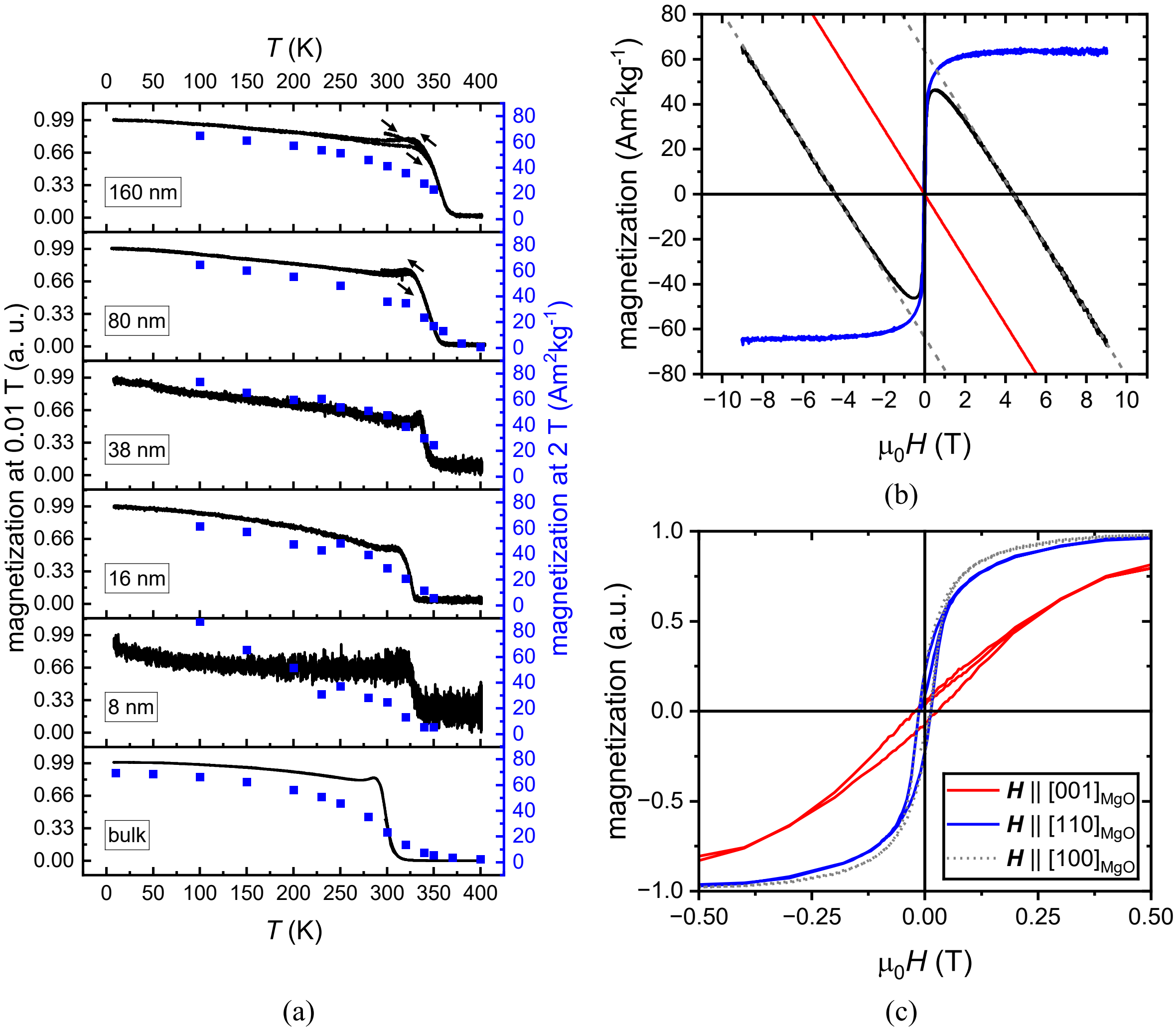

Fig. 3: (a) Thermomagnetic curves measured in-plane along $[100]_{\mathrm{MgO}}$ in 0.01 T magnetic field (black lines) and temperature dependence of magnetization at 2 T (blue squares). Heating and cooling are indicated by arrows when hysteresis is present. (b) Exemplary magnetization loop of 80 nm thick film at 100 K (black) decomposed into diamagnetic contribution from substrate (red) and ferromagnetic contribution from the film (blue). Linear extrapolation of saturated magnetization is indicated by dashed lines. (c) Normalized magnetization loop of 80 nm thick film at RT along $[110]_{\mathrm{MgO}}$ (solid blue line), $[100]_{\mathrm{MgO}}$ (dashed grey line), and $[001]_{\mathrm{MgO}}$ (solid red line).

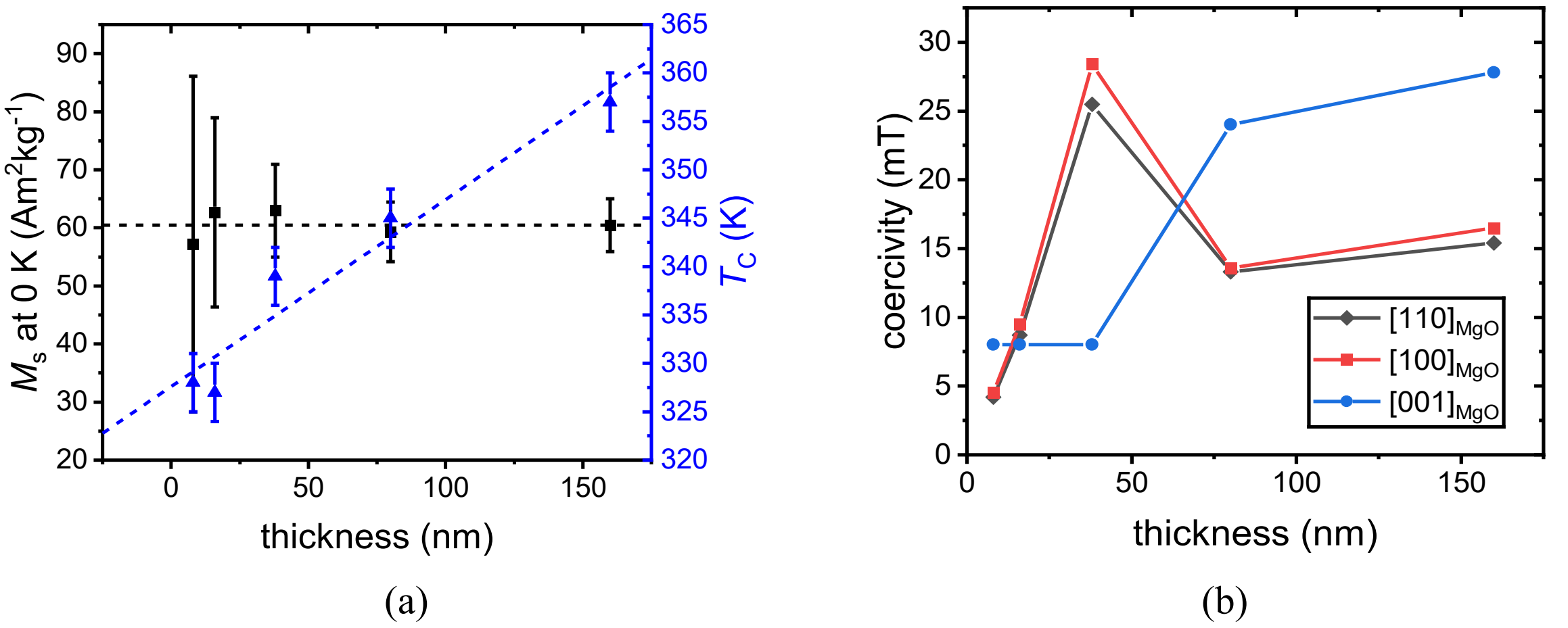

Fig. 4: (a) Thickness dependence of saturated magnetization $M_S(0)$ (black squares) and Curie temperature $T_C$ (blue triangles interlaced by dashed black line), fitted by constant and linear functions (dashed lines of corresponding colors). (b) Coercive fields for two in-plane directions, $[100]_{\mathrm{MgO}}$ (along the substrate edge) and $[110]_{\mathrm{MgO}}$ (45° to the substrate edge), and out-of-plane direction $[001]_{\mathrm{MgO}}$.

## C. Optical and magneto-optical properties

The spectra of $\varepsilon_{xx}$ resemble their single crystal counterparts [44,47,51,54,81] with classical metallic behavior dominated by intra-band contributions at lower energies and two inter-band electron transitions at around 1.6 and 3.2 eV (**Fig. 5 a**). These transitions are the most distinguishable in the Re $\sigma_{xx}$ spectra (dashed lines in **Fig. 5 a**) with two pronounced maxima at these energies. The $\varepsilon_{xx}$ spectra of the thinnest films (8 nm and 16 nm) significantly differ from each other and from the rest of the films, while in case of thicker films (38, 80, and 160 nm), they differ more subtly. The effect of the film-Cr interface might be responsible for these changes as the penetration depth for all films does not exceed 25 nm (see **Fig. S11 b** in the Supplementary Material [62]. Thickness dependences of $\varepsilon_{xx}$ for selected energies across the studied spectral range exhibit significant correlation to the in-plane coercivity (see **Fig. 5 b**). For Im $\varepsilon_{xx}$, we can see that apart from the spectral region 1.2–1.8 eV, in which the Pearson correlation coefficient (PCC) was changing the sign, PCC peaks in the vicinity of 3 eV (with the extrema of 0.96) and its absolute value remains above 0.5 (spectral behavior can be seen in **Fig. S13 b** in the Supplementary Material [62]). PCC for Re $\varepsilon_{xx}$ remains above 0.5 in the studied spectral range.

The experimental PMOKE spectra are shown in **Fig. 6**. The spectra exhibit similar spectral shape as in previous reports on single crystals [47,48,54]. Namely, PMOKE rotation (**Fig. 6 a**) has pronounced negative extrema around 1 and 3.8 eV with two smaller positive extrema in between (around 1.7 and 2.6 eV). Of the two smaller extrema, the one around 1.7 eV is higher in the martensitic phase, while, in the austenitic phase, the extremum around 2.6 eV is higher. This feature was also identified in single crystal spectra by Beran *et al.* [47]. While the lower energy extrema at 1 and 1.7 eV monotonically decrease with decreasing thickness, the higher energy extrema at 2.6 and 3.8 eV have the highest magnitude for the critical 38 nm thick film. Furthermore, significant spectral shifts of PMOKE rotation peaks located at 1.7, 2.6, and 3.8 eV were observed with varying thickness. With decreasing thickness, increasing photon energy seems to increase PMOKE extrema shifts, which rise from -0.2 eV for the low energy extremum at 1 eV to a staggering +0.7 eV for the 3.8 eV extremum. Similar trends are observable in the Kramers-Kronig related [82] PMOKE ellipticity spectra (**Fig. 6 b**). Namely, the shift of the 3 eV extremum by +0.5 eV with decreasing thickness and monotonic magnitude decrease for lower photon energies.

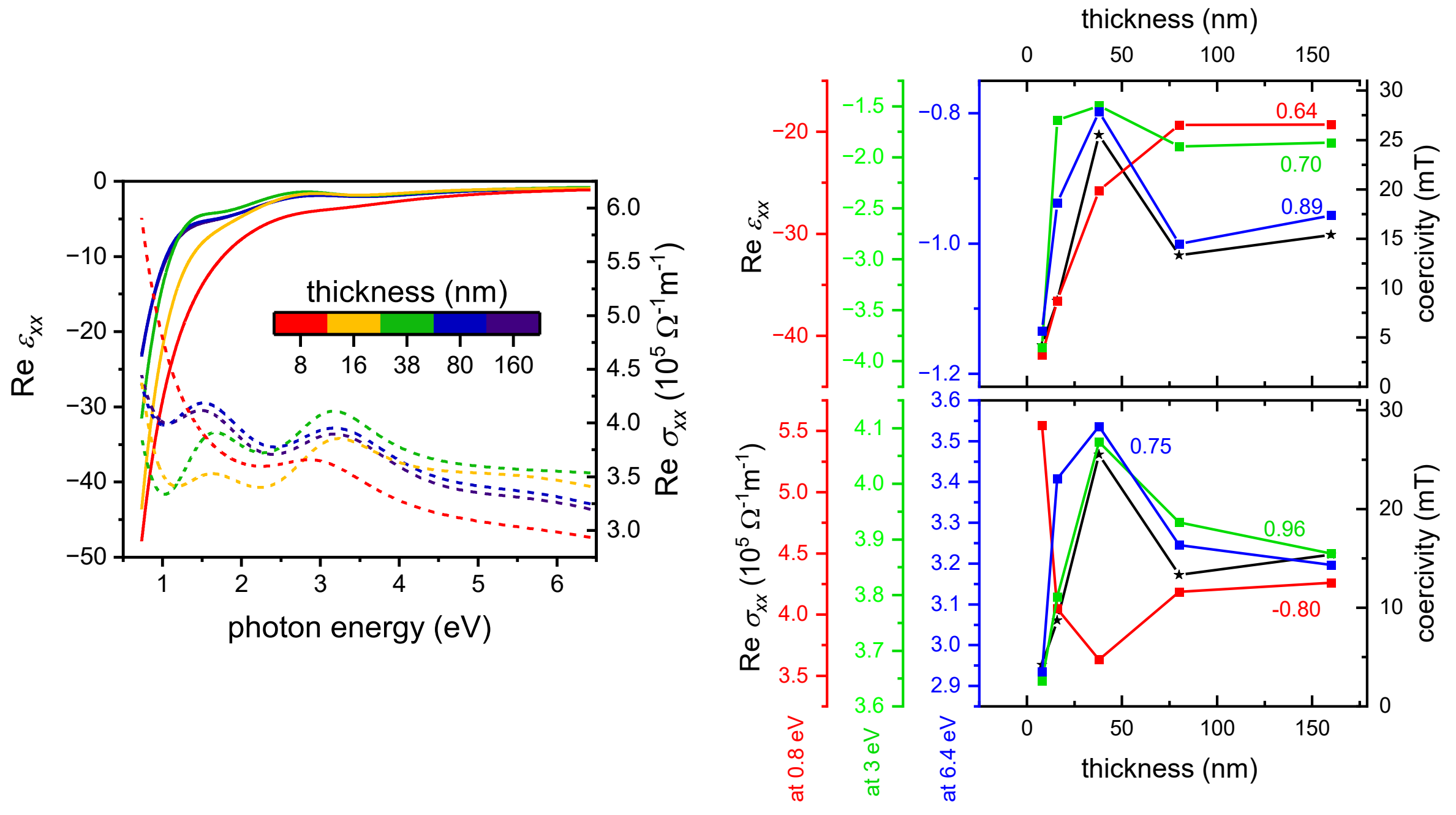

(a) (b)

Fig. 5: (a) Spectral dependence of Re $\varepsilon_{xx}$ (left axis, solid lines) and Re $\sigma_{xx}$ (right axis, dashed lines) for Ni-Mn-Ga films with thicknesses 8 nm to 160 nm. (b) Re $\varepsilon_{xx}$ (top part) and Re $\sigma_{xx}$ (bottom part) as functions of thickness at 0.8 eV (red), 3 eV (green), and 6.4 eV (blue). RT coercivity along the $[110]_{MgO}$ in-plane direction is plotted on right axes (black) and the numbers

found near each colored lines correspond to the Pearson correlation coefficient with the in-plane coercivity thickness dependence.

Combining the knowledge of the $\varepsilon_{xx}$ spectra as well as PMOKE spectra, we have deduced the spectral dependence of $\varepsilon_{xy}$. We have used the same model sample structure of thin Ni-Mn-Ga layer and Cr buffer layer on semi-infinite substrate as in the case of spectroscopic ellipsometry. We also considered and quantified the influence of possible oxide layer up to 2 nm thick using the Yeh formalism. Such oxide layer would primarily lead to variations comparable to the variation caused by the thickness error (±4 nm) and would have negligible spectral impact. Because this effect is below the instrumental resolution of our setup, we did not include oxide layer in the model.

Following the PMOKE trends, in the Re $\varepsilon_{xy}$ spectral extrema (**Fig. 7**) shifts with decreasing film thickness from 160 nm to 8 nm range approximately from -0.05 eV to more than 0.5 eV. Beginning at lower energy range, the 1 eV extremum seems to differ only between austenitic and martensitic phases with amplitude being larger in the former. Then, we see monotonic magnitude decrease with thickness decrease for the Re $\varepsilon_{xy}$ extremum at 1.6 eV or even more pronounced for the Im $\varepsilon_{xy}$ extremum at 1.9 eV. And finally, for higher photon energies, the thickness dependences of both parts are well correlated to the thickness dependence of coercivity (see **Fig. S13 b** in the Supplementary Material [62]).

The $\varepsilon_{xy}$ spectra of thin Ni-Mn-Ga films describe intrinsic material properties directly reflecting the electronic structure. To avoid losing useful information by overfitting the problem, the aim was to describe the $\varepsilon_{xy}$ spectra using the smallest viable number of semiclassical transitions, while still being able to reproduce the overall spectral shape. Therefore, we fit **Eq. (2)** to the data (**Fig. 7**) containing one paramagnetic transition located around 1 eV and two diamagnetic transitions around 1.8 eV and 3 eV. We indexed these transitions as 1, 2 and 3 by their increasing energy. In **Fig. 8**, one can see thickness dependences of the semiclassical transition parameters $A$, $E$, and $G$. Only the amplitude $A_2$ and the energy $E_2$ of the 1.8 eV diamagnetic transition showed no correlation. Most parameters were highly correlated to coercivity with the absolute value of PCC above 0.6. Notably, the 3 eV transition (blue) has its parameters (anti)correlated the most, with PCC $< -0.8$. Full correlation matrix can be found in **Fig. S14** in the Supplementary Material [62]. The only exception seems to be the amplitude $A_1$ of the 1 eV paramagnetic transition, whose absolute value might have been underestimated due to the relatively high noise in the IR region for the 8 nm thick film (see the bottom part of **Fig. 7**). It is possible that with lower noise, we would see positive correlation instead. Additionally, for the thinnest sample (8 nm), the agreement between the fitting function (dashed black line in **Fig. 7**) and the off-diagonal permittivity is evidently much lower than for other samples. Its thickness uncertainty is the largest (estimated up to 50 %, see **Table 2** or XRR data in the Supplementary Material [62]), which directly affects the calculated permittivity values. In the high energy region, transition 3 is very broad ($G_3$=2.6 eV) and shifted to 3.85 eV, while the Im $\varepsilon_{xy}$ spectral extremum is found at 3.45 eV. The reason for this discrepancy might originate either from lower detector sensitivity above 3.5 eV or there is additional high energy transition missing in the fitting function, which causes the minimization procedure to overestimate $E_3$ for the 8 nm thick film.

According to *ab initio* calculations [54,74,83–98], we expect transitions mainly between Ni 3$d$ and Mn 3$d$ states. Fuji *et al.* [83] proposed that the band Jahn-Teller effect stabilizes the martensite. Upon MT, a DOS peak exactly at $E_F$ would be split into two peaks, one above and one below $E_F$, decreasing the total energy. Indeed, *ab initio* calculations almost universally reported the splitting of Ni 3$d$ spin-minority states, regardless of whether the DOS peak in the austenitic phase was located precisely at or slightly below $E_F$. These predictions were corroborated by experimental data, including XPS [57] and polarized neutron scattering [99]. Other explanations of MT include strong electron-phonon interaction due to Fermi surface nesting [100–102] which result in softening of the $TA_2$ phonon mode at the nesting vector $\vec{Q} = \frac{1}{3}(1,1,0)\frac{2\pi}{a}$ as similar phonon anomalies had been previously reported in other SMA like Ni-Ti [103]. Notably, the *ab initio* studies of Nakata *et al.* [85], Entel *et al.* [89], Kart *et al.* [98], Gao *et al.* [91], Abbes *et al.* [94], and Obata *et al.* [97] predicted qualitatively similar DOS, which is further supported by our *ab initio* calculations of tetragonally distorted austenite (see the Supplementary Material [62]). For simplicity, we will limit our discussion to the density functional theory (DFT) results without the Hubbard $U$ parameter (its effect is investigated in Ref. [74]).

Below $E_F$ in the spin-minority channel, the Ni 3$d$ states spread from −3.5 to −1 eV, peaking around −1.5 eV, having by far the highest EDOS in this region. The spin-minority Ni 3$d$ $e_g$ states found at −0.2 eV are split in the martensitic phase by the band Jahn-Teller effect. Considering there are effectively no spin-majority states above $E_F$ to which electrons could be excited by photons with energy below 6 eV, only two DOS peaks in the spin-minority channel need to be considered. The first is approximately at 0.4 eV and the second is at 1.4 eV (in martensite, it is closer to 1.3 eV). The composition of the two peaks for both austenitic and martensitic phases is summarized in **Table 3**. Notably, the MT shifts the Ni $t_{2g}$ states from the higher energy peak to the lower energy peak. Thus, according to our semi-classical fit, the two diamagnetic transitions (at 1.8 and 2.8 eV in the martensitic phase) could be accounted for by excitation in the spin-minority channel from Ni $e_g$ states to the two peaks above $E_F$ as illustrated by **Table 3**. As for the 1 eV paramagnetic transition, the corresponding electron transition might take place between bonding and antibonding Ni $e_g$ states in the spin-minority channel near $E_F$. In martensite, taking spin-flip into consideration, the excitation might be happening from the highest energy spin-majority peak of Mn $t_{2g}$ and Ni $t_{2g}$ states at −0.8 eV to the 0.4 eV spin-majority DOS peak. However, in austenite, this spin-majority peak is even lower at −1.1 eV due to the contribution of Ni $e_g$ and Mn $e_g$ states, yielding energy 0.5 eV higher than what would be expected for the fitted paramagnetic transition.

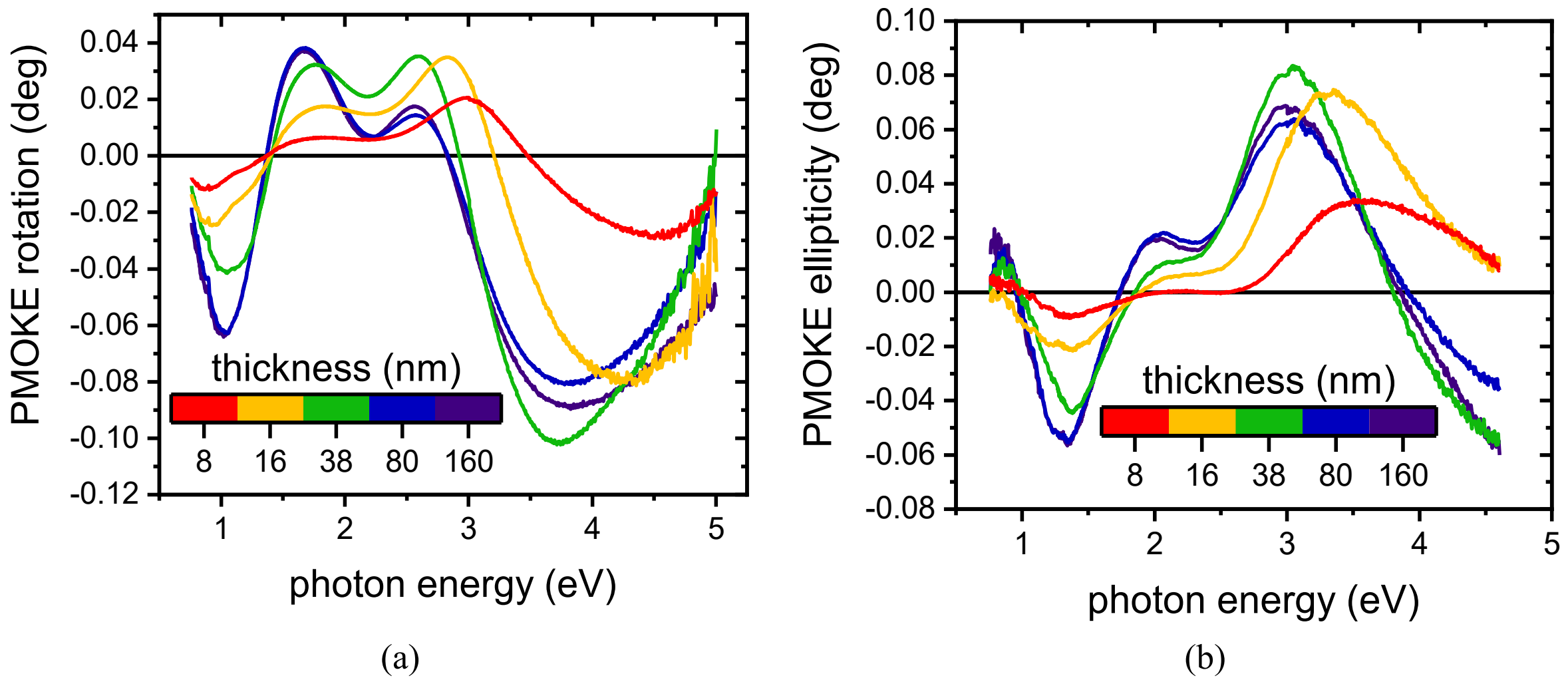

Fig. 6: PMOKE rotation (a) and ellipticity (b) spectra for 8, 16, 38, 80, and 160 nm thick Ni-Mn-Ga films.

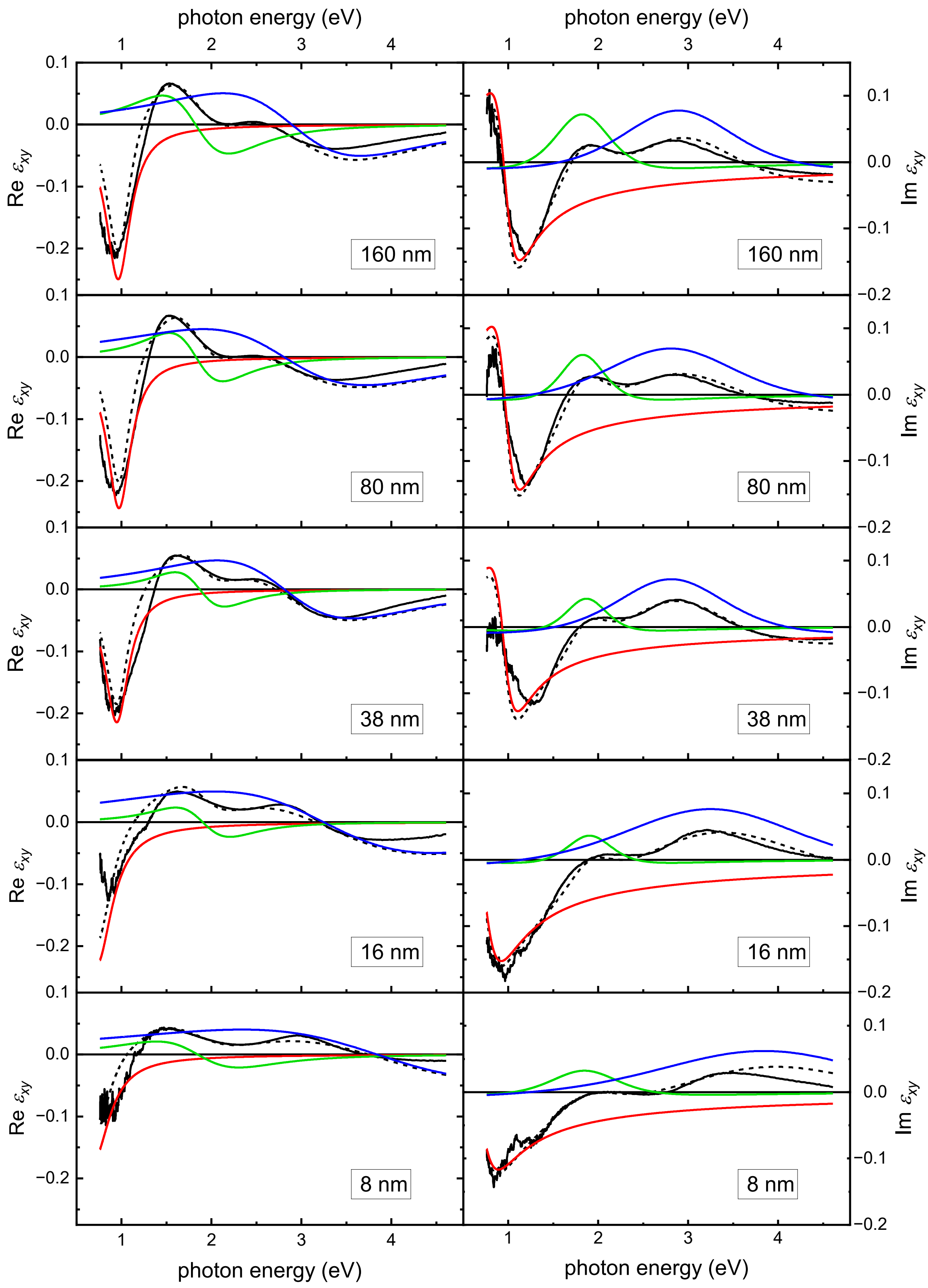

Fig. 7: The off-diagonal relative permittivity spectra (solid black lines) of Ni-Mn-Ga films with thicknesses from 8 nm to 160 nm (bottom to top row) with the real (imaginary) part plotted in the left (right) column. Semiclassical theory fit indicated by dashed black lines is decomposed into individual magneto-optical transitions. Namely, one paramagnetic (solid red lines) and two diamagnetic (solid green lines and solid blue lines) transitions.

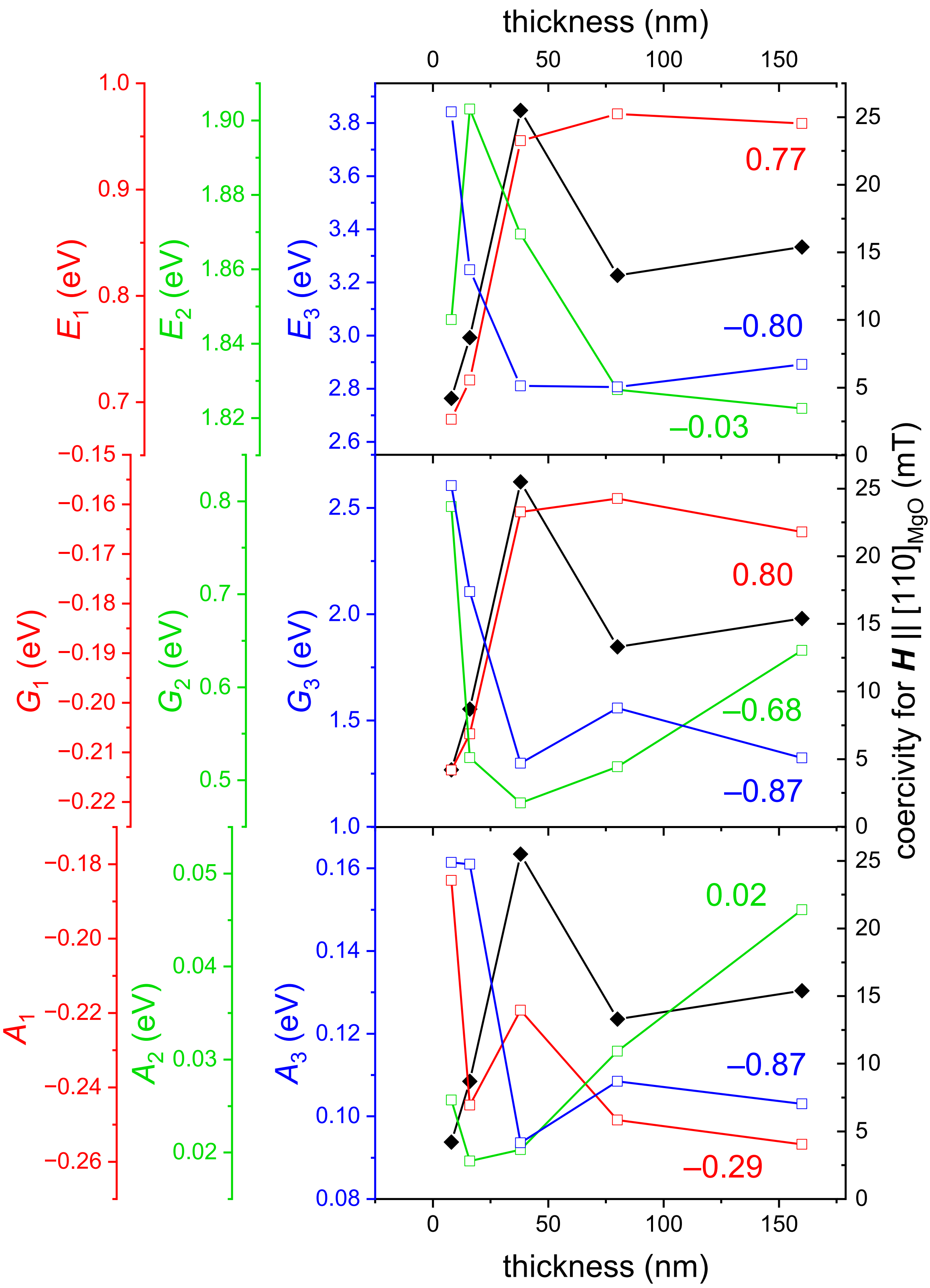

Fig. 8: Thickness dependences of energies $E_t$, widths $G_t$, and amplitudes $A_t$ of magneto-optical transitions (indexed by $t \in \{1, 2, 3\}$) obtained from semiclassical theory fit and in-plane coercivity (black) for applied field along $[110]_{MgO}$. One paramagnetic (red, $t = 1$) and two diamagnetic transitions (green and blue, $t \in \{2, 3\}$) were fitted to the $\varepsilon_{xy}$ spectra. The colored numbers found near each colored line correspond to the Pearson correlation coefficient quantifying correlation of thickness dependences between each magneto-optical transition parameter and in-plane coercivity (black).

Table 3: Summary of Mn and Ni $d$-states (ordered by decreasing DOS from top to bottom) in the spin-minority channel at 0.4 and 1.3 (1.4) eV above the Fermi level $E_F$ according to Entel *et al.*[73]. Note that Ni $t_{2g}$ states contribute to different peaks in each phase.

| phase | austenite | | martensite | |
|---|---|---|---|---|
| $E - E_F$ | 0.4 eV | 1.4 eV | 0.4 eV | 1.3 eV |
| Spin-minority $d$-states | Ni $e_g$ | Mn $e_g$ | **Ni $t_{2g}$** | Mn $e_g$ |
| | Mn $t_{2g}$ | Mn $t_{2g}$ | Mn $t_{2g}$ | Mn $t_{2g}$ |
| | | **Ni $t_{2g}$** | Ni $e_g$ | |

# IV. SUMMARY

This work presents a comprehensive study of the magneto-optical properties of epitaxial Ni-Mn-Ga films with thicknesses between 8 nm and 160 nm prepared by DC magnetron sputtering. Structural characterization revealed that thinner films (8–38 nm) retained the austenitic phase, while thicker films (80–160 nm) underwent martensitic transformation with a complex twin microstructure allowing MO comparison between differently strained films. XRD confirmed the presence of 14M and NM martensitic phases in thicker films, and AFM detected X-type and Y-type twin domains. In terms of magnetic properties, films thicker than 80 nm displayed martensitic transformation near the Curie point, whereas thinner films remained in the austenitic phase down to 10 K. In-plane coercivity increased with thickness in austenitic films due to gradual increase of stress, while martensitic films exhibited significantly higher out-of-plane coercivity due to complex twin microstructures.

Magneto-optical spectroscopy revealed great differences in electronic structure between martensite and austenite phases in thin epitaxial films. Using semiclassical theory of magneto-optical transitions, three inter-band magneto-optically active transitions between Mn and Ni $d$-states were identified in the permittivity spectra. Fitted parameters of these transitions were shown to be influenced by the substrate-induced strain as apparent from significant correlation to the thickness dependence of coercivity.

This study provides valuable insights into the relationship between film thickness, electronic structure, and magneto-optical properties in epitaxial Ni-Mn-Ga films.

# V. ACKNOWLEDGEMENTS

This work was supported by Czech Science Foundation Grant No. 24-11361S, OPJAK – Ferroic multifunctionalities (FerrMion) CZ.02.01.01/00/22_008/0004591, the Ministry of Education, Youth and Sports of the Czech Republic through the e-INFRA CZ (ID:90254), and the Charles University grant SVV–2024–260720.

Authors gratefully acknowledge D. Musiienko for assistance with XRF measurements, L. Fekete for support with AFM characterization, and T. Kmječ for TEM imaging. Their contributions were essential to the completion of this work.

Experiments were performed in MGML (mgml.eu), which is supported within the program of Czech Research Infrastructures (project no. LM2023065).

During the preparation of this work the authors used ChatGPT (OpenAI) to improve the readability of the text. After using this tool, the authors reviewed and edited the content as needed.

# VI. DATA AVAILABILITY STATEMENT

The data that support the findings of this article are openly available [104].